\documentclass{amsart}
\usepackage{amsfonts}

\newcommand{\myel}{{\mathcal L}}

\newcommand{\nei}{{\mathcal N}}

\newcommand{\amin}{\arg\min}

\newcommand{\gee}{{\mathcal G}}
\newcommand{\dee}{{\mathcal D}}
\newcommand{\eps}{\epsilon}

\newcommand{\cspace}{\mbox{, }}

\newcommand{\other}{\mbox{\rm {otherwise}}}

\newcommand{\te}{{\tau(\eps)}}
\newcommand{\X}{{\mathcal X}}
\newcommand{\finv}{{f^{-1}}}

\newcommand{\minf}{\left( \left. \left[M, \infty \right. \right) \right)}

\newcommand{\avmomgen}{{\ee^\ast \left[ e^{ \xi \te} \right]}}

\newcommand{\ejx}{{e^{j\xi}}}

\newcommand{\pjejx}{{p^j \ejx}}

\newcommand{\peemu}{{{\rm P}^\ast}}

\newcommand{\sumj}{\sum_{j=0}}

\input{epsf}

\theoremstyle{definition}

\theoremstyle{remark}

\numberwithin{equation}{section}

\newcommand{\ee}{{\bf E}}

\begin{document}

\title{Robustness and Evolvability of the B Cell Mutator Mechanism}

\author{Patricia Theodosopoulos}
\author{Ted Theodosopoulos}
\address{Department of Mathematics and \\ Department of Decision Sciences
\\ Drexel University \\ Philadelphia, PA 19066}
\email{theo@drexel.edu}
\urladdr{faculty.lebow.drexel.edu/TheodosopoulosT}


\date{May 14, 2005.}



\begin{abstract}
We present a model that considers the maturation of the antibody population following primary antigen presentation as a global optimization problem \cite{tedpatty}.  The trade-off that emerges from our model describes the balance between the safety of mutations that lead to {\it local} improvements in 
affinity and the necessity of the system to undergo 
{\it global} reconfigurations in the antibody's shape in order to achieve its 
goals, in this example of fast-paced evolution.  The parameter $p$ which quantifies this trade-off appears to be itself both robust and evolvable. 
This parallels the rapidity and consistency of the optimization operating during the biologic response.  In this paper, we explore the robust qualities and evolvability of this tunable control parameter, $p$.
\end{abstract}

\maketitle

\section{Introduction}

The challenge that faces the immune system is the combinatorial complexity of antigenic agents and the need to reliably, rapidly and consistently respond with the available repertoire.  This entails a considerable expense of energy to the organism and risk, in the form of potentially creating a malignant or autoimmune state.  
These qualities make affinity maturation well suited to an approach through mathematical modeling.  
The purpose of our model is to study the phenomenon of affinity maturation of the humoral immune response as a global optimization problem on a rugged affinity landscape. Affinity in our model is defined as the probability of antigen and antibody existing in the bound state. Movement on the affinity landscape occurs through mutations in the immunoglobulin (Ig) gene that produce changes in configuration of the Ig molecule.  

This modeling approach aims not only to replicate some of the observed behaviors of the system, but also 
to suggest a more fundamental property of the system that might be underpinning the observed behavior.  
This property is embodied in a control parameter that the model utilizes to explore the landscape and produce a consistent and adequate response in a limited period of time.  

This parameter is intended to capture the trade-off between local steps on the landscape of affinity that produce small structural changes and potentially produce only incremental increases in affinity, versus 
global jumps that might produce large changes in antigen combining site configuration.  
These global jumps, although risky, allow the system to avoid becoming trapped in local optima and may 
permit faster exploration of the landscape and faster attainment of a higher affinity.  

The model system exhibits the qualities observed in the biologic system, which include a fast and robust response optimization despite variability in the landscape structure which reflect
novel antigenic challenges.  The model system also allows for the inclusion of other observed behaviors of the biologic system such as the repertoire shift \cite{fukita,klix,diaz,blanden}. The presumed instantiation of the trade-off parameter is in the 
genetic code, thus making it subject to the forces of evolution and selection as well. 
This paper begins with a brief review of the dynamics of the model and the evolutionary landscape.  
We subsequently focus on the importance of $p$, and its relation to performance robustness, and the affinity threshold for terminating the hypermutation process.  We also attempt to clarify the observed evolution of $p$.

\section{The Model}

The model attempts to delineate a selection-based 
trade-off that underlies the evolution of affinity-matured antibodies.  
We take into consideration both the contribution of the microenvironment in the germinal center as well 
as the intrinsic properties of antigen--
antibody interactions.  The main purpose of the model is to 
elucidate the drivers behind the apparently complex behavior of 
the immune system and develop a qualitative understanding of what 
makes the system work so efficiently. The model does not attempt to represent 
the detailed molecular complexity of the system.  Rather, we wish to model the 
qualitative aspects of the system that lead to the observed large scale 
behaviors of the system.  Through this exercise, we hope to gain an appreciation and understanding of the more fundamental mechanisms that drive the system so effectively towards its goal.

\subsection{Local Steps versus Global Jumps}

The model we present attempts to capture the biological ``trade-off'' that 
controls the mutation process and enables the rapid generation 
of high affinity antibodies. One might understand this trade-off 
in terms of the balance between mutations that produce only local changes in the 
conformation and are therefore more likely, although not 
exclusively, to lead to incremental changes in the affinity, 
versus those mutations that produce global {\it jumps} in shape space and 
therefore enable the system to avoid becoming trapped in local optima of the affinity landscape. 

The na\"{\i}ve repertoire usually produces antibodies with affinities on the order of $10^5 M^{-1}$ and the hypermutation process results in antibodies with a range of affinities from $10^6 - 10^8 M^{-1}$ within a period of days (corresponding to around $10 - 20$ generations) from a finite number of clones \cite{elgert}. 
The traditional treatment of affinity changes and 
mutational studies favors the idea that the mutation 
process enables the selection of clones that undergo a stepwise 
increase in affinity \cite{braden,nossal,brown} -- an additive 
effect of changes that create new hydrogen bonds or new
electrostatic or hydrophobic interactions between the residues of 
the antigen epitope and the antibody variable region, and can alter these interactions with 
associated solvent molecules \cite{covell,andersson,braden}.  However, it is observed that all codon changes 
cannot be translated into stepwise energetic changes 
\cite{covell}. In the literature, affinity increases are sometimes understood as improved kinetics, often 
translated into a lower $K_{\rm off}$ or a higher 
$K_{\rm on}$, although which kinetic component dominates during different phases of the 
hypermutation process is unclear \cite{wedemayer,andersson,nayak,batista,foote2}. Other mutations may be neutral from an affinity perspective, but may actually be permissive of subsequent affinity enhancing mutations \cite{cowell,furukawa,huynen}. 
	
Models of affinity changes as stepwise energetic improvements in the selected antibodies have also led to the idea 
that the antibody conformation evolves likewise. The 
progression to a higher affinity conformation conformation occurs at the 
expense of entropy, in exchange for a decrease in enthalpy and a 
commensurate increase in affinity \cite{wedemayer}. Since we 
cannot reliably observe the process, we cannot presume that this 
stepwise search is what is always functioning in the germinal 
center.  
Our modeling paradigm predicts that the observed rapid elaboration of high affinity 
antibodies through the germinal center reaction could only occur if the system experiences occasional large {\it jumps} in order to 
more efficiently sample the conformational landscape.

It is likely that the optimization takes 
advantage of bias in the V gene code and that subsequent mutations would attempt to create flexibility in 
some regions to facilitate docking while other regions are optimized to maximize antigen-antibody interactions and 
stabilize binding for appropriate feedback and signaling to occur \cite{furukawa,guermonprez}.
The positions of positively selected mutations show that 
replacement mutations occur preferentially in the complementarity determining regions (CDRs) versus the intervening 
framework regions (FRW)\cite{foster2,foster}. The FRW is often described as being very sensitive to replacement mutations, but it 
appears now that they too can tolerate a certain number of 
replacement mutations, and that the CDRs may alternately possess 
a sensitivity to mutations through the coding structural elements 
\cite{wiens}. The greatest diversity is seen in 
CDR 3, which typically has the most contact residues with the 
antigen, while CDR 1 and 2 usually comprise the sides 
of the binding pocket \cite{schroeder}.  

Studies comparing germline diversity with 
hypermutated V genes show that the amino acid differences 
introduced by mutation were fewer than the underlying diversity of 
the primary repertoire. These studies further suggest that the CDR 1 and 2 residues, which are more conserved in the na\"{\i}ve repertoire and
often create the periphery of the binding site, are favored for 
mutation. Alternatively, the more diverse residues represented in the na\"{\i}ve repertoire are generally not as favored for 
mutation \cite{dorner}. Perhaps there can even be identified critical residues where mutation might produce large conformational changes.

Although certain point mutations in critical residues may create large changes in conformation, additional mechanisms for global 
jumps may be appreciated from evidence of receptor editing \cite{dewildt} in the germinal center as well as 
the relative frequency of deletions and insertions.
\cite{klein}.  These types of alterations would also be expected to 
represent {\it global jumps} on the affinity landscape.

\subsection{The Evolutionary Landscape}

As mentioned earlier, we treat the affinity
maturation of the primary humoral immune response as a problem of
global optimization.  This paradigm should be contrasted with the
``population dynamics'' modeling approach.  The latter class of models entails
the tallying of individual immune cell types and the investigation of
the transition dynamics between their allowable states.  Such models
represent the emergence of affinity optimization as a result of these
cell population dynamics.  In this vein, it is generally the evolution
of the average affinity in the population that is the dominant
variable.

In our model the primary role is played by the order statistics (e.g. maximum) of the affinity levels that have been achieved up to 
any stage of the hypermutation process.   Specifically, the size of the B cell population is exogenous to our model.  
We assume that the hypermutation process is initiated via a
mechanism outside the scope of our model.  Our treatment of the
hypermutation process terminates upon the development of a desirable proportion of clones with
sufficiently high affinity.  As a result of our focus on the
affinity improvement steps, we measure time in a discrete fashion by
counting inter-mutation periods.

As a first step, we begin with the space of all DNA 
sequences encoding the variable regions of the Ig molecules and a 
function on that space that models the likelihood that the 
resulting Ig molecule becomes attached to a particular antigen.  
This {\it affinity function} is conceptualized in a series of 
mappings which portray the biochemical mechanisms involved.  

To begin with, the gene in question is transcribed into RNA and 
subsequently translated into the primary Ig sequence.  This step 
describes the mapping from the genotype (a 4--letter alphabet per 
site) to the sequence of amino acids making up the Ig molecule (a 20--letter alphabet per site).  
The next step is the folding of the 
resulting protein into its {\it ground state} in the presence of 
the antigen under consideration.  This step is modeled as a 
mapping from the space of amino acid sequences to the three-
dimensional geometry of the resulting Ig molecule\footnote{This 
concept is analogous to that of {\it shape space} in Chap. 13 
of \cite{rowe}, first introduced in \cite{perelson2} and further elaborated in \cite{perelson3}.}.

Finally, the protein shape gives rise to the free energy of the Ig 
molecule in the presence of the antigen.  The free energy in turn 
is used to define the association/dissociation constants and the 
corresponding Gibbs measure which determines the likelihood of attachment.  The 
resulting affinity is visualized as a high-dimensional landscape, 
where the peaks represent DNA sequences that encode Ig molecules 
with high affinity to the particular antigen.

Let $\X$ denote the space of DNA sequences encoding the $V_H$ and $V_L$ 
regions of an Ig molecule.  Let $f$ be a positive, real-valued 
function on $\X$, which denotes the affinity function, as 
described above.  Finally, consider the gradient operator $\dee 
f(x) = \amin_{y \in \nei(x)} f(y)$, where $\{\nei(x) \subseteq \X 
\cspace \/ x \in \X\}$ describes the neighborhood 
structure\footnote{In this paper we concentrate on point mutations 
as the mechanism for local steps and thus the neighborhood we 
consider consists of all 1--mutant sequences.  It has been 
suggested \cite{manser2,goyenechea} that more than one 
point mutation may occur before the resulting Ig molecule is 
tested against an antigen presenting cell to determine its 
affinity.  Our model can capture such an eventuality by 
appropriately modifying the neighborhood structure to include the 
2-- or generally $k$--mutant sequences.} in $\X$.  With this notation, 
for each sequence $x \in \X$, successive applications of the 
gradient operator converge to the closest local optimum, i.e., 
there is a finite positive integer $d(x)$ and a genotype 
${\mathcal F}^\ast (x) \in \X$ such that for all $n \geq d(x)$,
$$\dee^n f(x) = \dee^{d(x)} f(x) = {\mathcal F}^\ast (x).$$
This association partitions $\X$ into subsets that map to 
the same integer under $d(\cdot)$.  These ``level sets'' contain all 
sequences that are a fixed number of point mutations away from 
their closest local optimum.  

A further ingredient of our model for the affinity landscape is 
the relative nature of the separation between strictly local and 
global optima.  In practice, the global optimum is not necessarily 
the goal.  Instead, some sufficiently high level of affinity is desired.  
This affinity threshold is generally unknown a priori.  Our model 
allows us to view the landscape as a function of the desired 
affinity threshold.  We are able to 
study the dependence of our model's performance for a variety of affinity thresholds and thus investigate the trade-off between the desired affinity and the required time.

The appreciation of strictly local versus global optima as a 
relative characteristic necessitates a finer partition of the 
level sets.  Specifically, for each level of affinity threshold, 
some of the local optima in $\X$ are below it and therefore are 
considered strictly local, while others are above it and are 
therefore considered global.  This leads to a decomposition 
of each level set into the part containing sequences a certain 
number of steps below a strictly local optimum versus sequences 
whose closest local optimum is also global because its affinity is 
above the desired threshold.

\subsection{Optimization Dynamics}

We model the dynamics of evolutionary optimization on the affinity 
landscape as a Markov chain.  Specifically, the chain may take one 
of two actions at each time step:  it may search locally to find 
the gradient direction, and take one step in that direction or it 
may perform a global jump, which effectively randomizes the chain.  
The decision between the two available actions is taken based on a 
Bernoulli trial with probability $p$: when $p=0$, the chain 
performs global jumps all the time while $p=1$ prohibits any 
global jumps.  Thus, the parameter $p$ controls the degree of 
randomization in the Markov chain.  The biological distinction 
between local search versus global jumps is realized by means of 
at least three mechanisms described earlier: receptor editing, deletions/insertions 
and point mutations that lead to sizeable movements in shape space. 

The mathematical description of the Markov chain model described 
above uses the following generator:
$$\left[ \gee (p) \phi \right] (x) \stackrel{\Delta}{=} p \phi(\dee 
f(x)) + (1-p) {\bf E}^{\mu} [\phi] - \phi(x),$$
where $\mu$ represents a global mixing measure, e.g. the uniform measure with support spans the entire relevant sequence space.  Note that we suppress the dependence on the level of the control parameter $p$ (e.g. writing $\gee$ instead of $\gee_p$) for notational clarity and convenience.  Keep in mind though that all variables associated with this Markov Chain will, implicitly, be functions of $p$.
We are interested in estimating the extreme left tail of the 
distribution of the exit times for the resulting Markov chain.  In 
particular, let
$$\tau(M) \stackrel{\Delta}{=} \inf \left\{k \geq 0 | X_k \in 
\finv \minf \right\},$$
where $X_k$ denotes the Markov chain under consideration.  We are 
interested in estimating the likelihood that at least one out of a 
population of $n$ identical, non-interacting replicas of the 
Markov chain will reach an affinity level higher than $M$ before 
time $y$.  It should be noted that, by virtue of the discrete 
nature of the Markov chain, time in this context is measured by 
the number of mutation cycles experienced by the system.  The 
probability we are looking for takes the form
$$\peemu \left(\tau_{\left(n, \lceil qn \rceil \right)} (M) \leq y \right) = 1- 
\sum_{i=0}^{\lfloor qn \rfloor} \left( \begin{array}{c} n \\ i \end{array} \right) \peemu \left(\tau_1 (M) \leq y \right)^i \left( 1- \peemu \left(\tau_1 (M) \leq y \right) \right)^{n-1},$$
where $\peemu$  denotes the path measure induced by the Markov 
chain the index $i$ tallies the replica under consideration, and we use the notation $X_{(n,m)}$ to denote the $k$th order statistic out of a sample of $n$ independent draws from the distribution of the random variable $X$.  
Since we are focusing our attention to the GC reaction, $n$ is 
approximately $10^3$-$10^5$.

\subsection{Methodology}

\begin{figure}
\epsfxsize=5in
\epsfbox{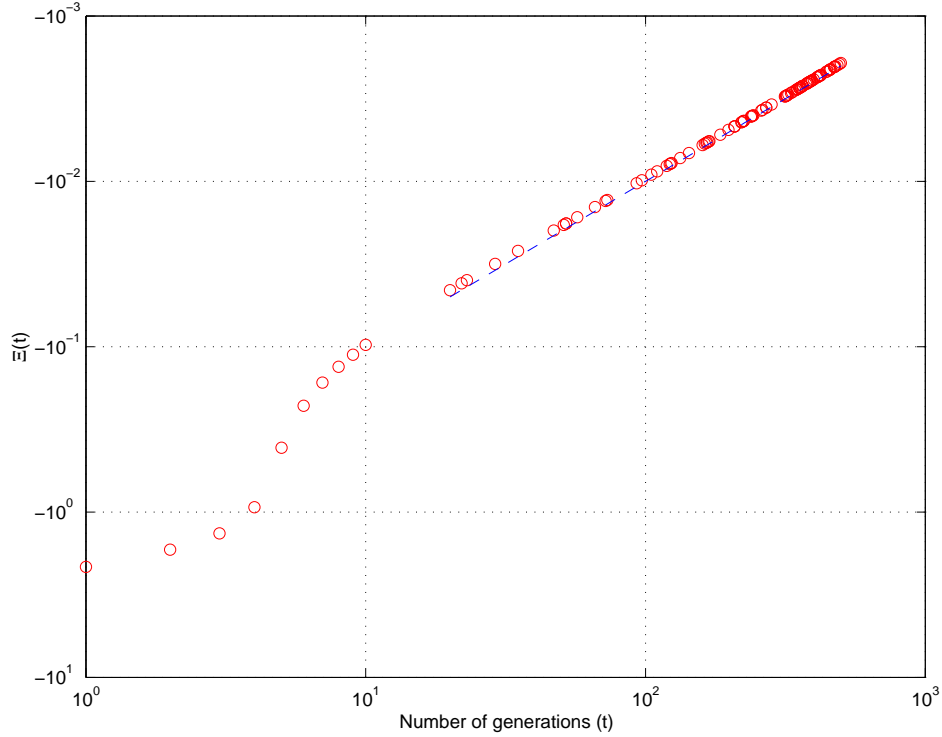}
\caption{Approximation scheme for $\Xi (t)$ as $t \searrow 0^+$ (incidentally, for this example, $\ee^\ast [\tau] = 16,473$)}
\label{fig:approx1}
\end{figure}

The study of the evolutionary optimization process outlined in the 
previous section uses results by the second author on the 
covergence rates of exit times of Markov chains \cite{ted1a,ted3}.  
The general approach for estimating the desired 
tails of the exit time distributions consists of the following 
steps:

\begin{itemize}
\item [{(i)}] 
We formulate a Dirichlet problem for $\gee (p)$ on $\finv \minf$ whose 
solution provides a martingale representation of the Laplace 
transform of the exit time $\tau(M)$.

\item [{(ii)}] 
We solve the resulting Dirichlet problem and compute the desired 
Laplace transform as
$$\psi(\xi) \stackrel{\Delta}{=} \avmomgen = {\frac {\left(1- 
pe^\xi \right) \sumj^b q(j) \pjejx}{1- e^\xi +
(1-p) e^\xi \sumj^b q(j) \pjejx}}$$
where $\ee^\ast$ denotes the expectation under $\peemu$ starting from a 
$\mu$--distributed initial sequence.  Here $q(\cdot)$ denotes the measure of the level sets a given number of local steps below a global optimum (i.e. an optimum above the desired affinity threshold), as described in the section on the Evolutionary Landscape.

\item [{(iii)}] 
We compute the Legendre-Fenchel transform $I(y)$ of the cumulant 
of $\tau(M)$ as
$$I(y) = \left\{ \begin{array}{ll}
\int_{1}^{\frac {y}{\ee^\ast [\tau]}} \Xi(t) dt \cspace & \mbox{if 
$y \geq \ee^\ast [\tau]$}
\\ 
\int^{1}_{\frac {y}{\ee^\ast [\tau]}} \Xi(t) dt \cspace & \other
\end{array} \right. ,$$
where $\Xi(t)$ is the (positive or negative depending on whether 
$y \geq \ee^\ast [\tau]$ or not) solution to 
$${\frac {d \psi}{d \xi}} \left( {\frac {\Xi(t)}{ \ee^\ast 
[\tau]}} \right) =t\ee^\ast [\tau] \psi \left( {\frac 
{\Xi (t)}{ \ee^\ast [\tau]}} \right).$$
It turns out \cite{ted1a} that, for $y \leq \ee^\ast 
[\tau]$,
$$\peemu \left(\tau_i (M) \leq y \right) \cong \exp \left\{ I(y) 
\right\}.$$
Note that $\ee^\ast [\tau]$ can be evaluated explicitly using the representation of the Laplace transform.  In particular, one can show after some algebra that the expected number of generations required for a particular antibody to evolve above-threshold affinity is given by:
$$\ee^\ast [\tau] = {\frac {1}{(1-p) \sumj^b q(j) p^j}},$$
which, for typical parameter values, tends to be over 10,000!  What we are after is the extreme left tail of the response time distribution, the few antibodies in the GC population that happen to take significantly fewer generations than average (typically between $10$ and $20$) to evolve above-threshold affinities.

\item [{(iv)}] 
We estimate $\Xi(t)$ by performing a Taylor expansion of the 
cumulant of $\tau(M)$ at $-\infty$, yielding
$${\frac {d \psi}{d \xi}} \left(\log \lambda \right) 
=\sum_{i=1}^\infty {\frac {c_i \lambda^i}{i!}},$$
as $\lambda \searrow 0^+$.  Inverting the polynomial on the right-hand side we obtain the general approximation form 
$$\Xi(t) \cong -t^{-1},$$
which holds remarkably well as long $t \geq 10$, i.e. at least ten generations.  For $t=1,2, \ldots , 9$ we evaluate $\Xi(t)$ numerically, by solving the associated polynomial.  Figure \ref{fig:approx1} shows an example of this approximation scheme.
\end{itemize}

\subsection{Note on random variables}

Three distinct sources of randomization play a central role in our model. In an effort to allay confusion, we want to be particularly clear about our use of random variables in the description of our results. 

Firstly, there is a dimension of randomness arising fundamentally from the random nature of the mutations during the immune maturation process.  We refer to different instances of this process as ``mutation scenarios'' and usually we reserve the letter $\omega$ to signify them.  The parameter $p$ introduced earlier serves to control this dimension of randomness.  

Secondly, there is variability within the population of antibodies that make up a particular germinal center under investigation.  We will generally only be concerned with the proportion of the antibody population from a given germinal center that shares a common property, rather than the explicit identification of any one antibody out of the population.  Thus, we use the letter $q$ to denote proportions of this population, rather than denoting random elements explicitly. 

Thirdly, there is a dimension of randomness arising from the unknown genotype of the antigenic challenge.  As described above, each antigen generates a different affinity landscape within which the antibody evolution process occurs.  We typically reserve the letter $z$ to denote a random instance of an affinity landscape (resulting from a random antigen).

As will become apparent in the following sections, our results typically involve properties of appropriately chosen quantiles of the ``response distribution'', i.e. the random variable measuring the time (in mutation cycles) until a given proportion of the antibodies populating the germinal center under investigation reaches an affinity for the antigen under investigation above a given threshold.  Mathematically, these are quantiles of a stopping time under the probability measure of $\omega$.  As such, these objects are still random variables, relative to the choice of a random antigen and resulting affinity landscape $z$.  Later we proceed to describe properties of the response distribution relative to randomly chosen antigens.  Mathematically, these are quantiles with respect to the probability measure of $z$.  

We generally use analytic estimates to study the properties of the distribution of $\omega$ while reserving simulation techniques for the properties of the distribution of $z$.  For instance, Figures \ref{fig:prop1} and \ref{fig:prop2} are generated analytically, while Figures \ref{fig:meanstd1}, \ref{fig:cumrobust1} and \ref{fig:cumrobust2} are generated by repeating the analytical estimates for 1,000 randomly generated affinity landscapes.

\section{Overview of Results}

The setting described above was used in \cite{tedpatty} to obtain estimates on the response characteristics of the immune system and its sensitivity to the amount of mixing in the MC.  More concretely, let $\myel(a,b,c)$ denote the set of affinity landscapes, parameterized by 

\begin{itemize}
\item the maximum number of affinity level sets $a$ in a strictly local maximum hill,  
\item the maximum number of affinity level sets $b$ in a global maximum hill, and
\item the ratio $c$ of the size of the affinity level sets that belong to the attraction basin of a strictly local maximum versus those that belong to the global maximum attraction basin.
\end{itemize}

Also, let $\tau \left(\omega, z, p, q \right)$ denote the random number of mutations that it takes under mutation scenario $\omega$ for a proportion $q$ of the population of antibodies in a particular GC to reach the desired level set in the random affinity landscape $z \in \myel(a,b,c)$ for some $a,b \mbox{ and } c$, while performing global jumps with probability $1-p$.  This is a generalization of the stopping time $\tau (M)$ defined earlier in the section on Optimization Dynamics.  We now suppress the dependence of the stopping time on the affinity threshold $M$, focusing instead on the other variables that influence it.  We can summarize the results of \cite{tedpatty} as follows:

\begin{enumerate}
\item For a population $N$ of a few tens of thousands and biologically justifiable values of $a,b \mbox{ and } c$ (see for example \cite{elgert,manser2,ramit}), for any $p$ above some minimum value and for all $z$, ${\rm Pr} \left(\tau \left(\cdot, z, p, N^{-1} \right) \leq y \right)$ exhibits a sharp cutoff \cite{oprea}
\item Let $R \left(c_1, z, p, q \right)$ denote the $c_1$-quantile of the distribution of $\tau$.  Then, for any choice of $c_1$ and for all $z$, $R \left(c_1, z, \cdot, N^{-1} \right)$ exhibits a unique minimum at a $p^\ast < 1$.
\item The graph of the mean and standard deviation of $R \left(c_1, \cdot, p, N^{-1} \right)$ over randomly generated landscapes has two branches, one for low $p$ and one for high $p$, that we designate ``liquid'' and ``solid'' state respectively.
\item Let us begin the MC with a randomly chosen value of $p$ and allow $p$ to evolve taking incremental steps after each antigenic response, with a bias towards the direction of the optimal value of $p$ for the most-recently resolved antigenic affinity landscape.  Under these conditions, the value of $p$ evolves autonomously to a narrow range (strictly less than 1) around the optimal values of the majority of affinity landscapes (``optimal range'').
\end{enumerate}

What we aim in the current paper is to explore three of these areas in more detail.  Specifically, we first investigate how the cutoff behavior changes when we require that $q>N^{-1}$.  We then attempt to better describe the characteristics of our conjectured phase transition when we contrast the responses of high-$p$ vs. low-$p$ MCs across randomly generated landscapes.  Finally, we define a more realistic version of the evolving $p$ model and show that the system continues to converge to a value of $p$ in the optimal range.

\section{Evolving Population of Antibodies}

\begin{figure}
\epsfxsize=5in
\epsfbox{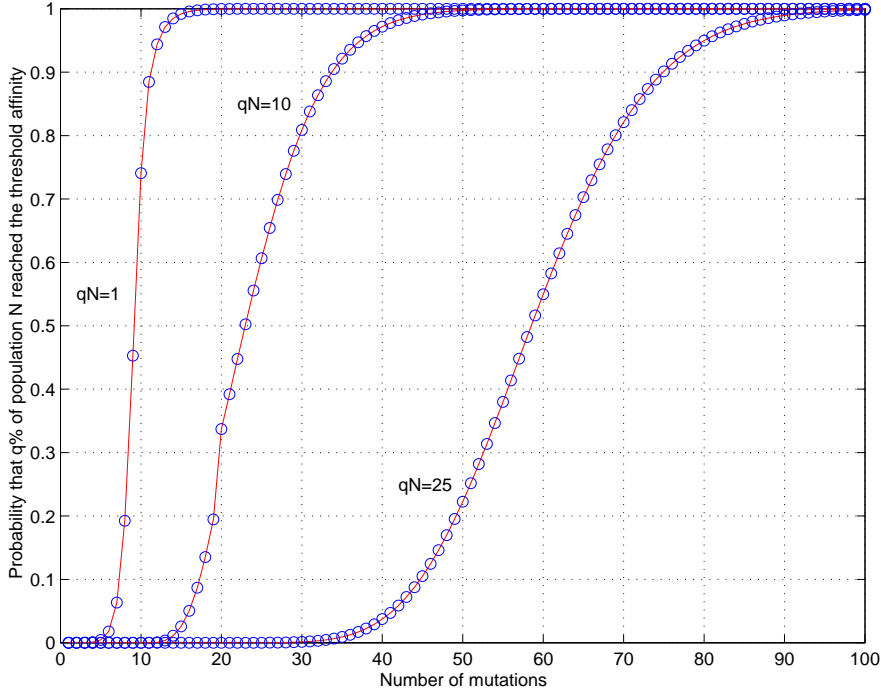}
\caption{Convergence of ${\rm Pr} \left(\tau \left(\cdot, z, p, q \right) \leq y \right)$ as a function of $y$ for three different values of $q, p=0.8, a=20, b=10 \mbox{ and } c=1,000$}
\label{fig:prop1}
\end{figure}

We begin by studying the propagation of affinity maturation across the population of antibodies in a GC.  Figure \ref{fig:prop1} displays the cumulative probability that a proportion $q$ of the $N$ antibodies populating the GC have reached the desired affinity threshold after an increasing number of accumulated mutations.  The cutoff observed in \cite{tedpatty} for the case $q=N^{-1}$ (requiring a mere five mutations to boost the probability that at least one antibody in the GC has reached the threshold affinity from below 10\% to over 90\%) becomes substantially attenuated as we increase $q$.  Already when we ask that at least 25 different mutant antibodies, out of the approximately 50,000 assumed to populate the GC, reach the affinity threshold, the sharp cutoff has been replaced by a gradual transition that requires more than 30 mutations to affect the same increase in cumulative probability.  

\begin{figure}
\epsfxsize=5in
\epsfbox{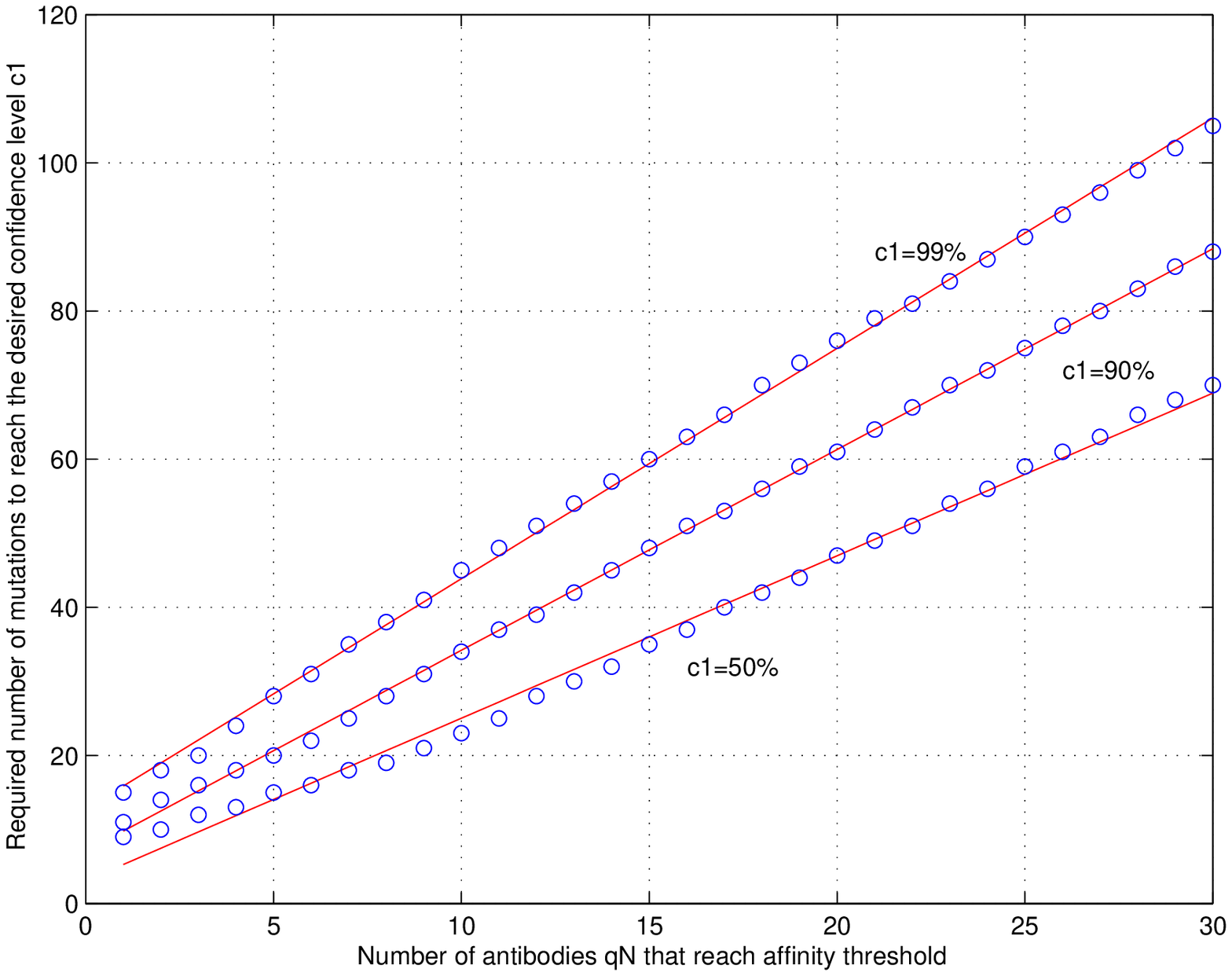}
\caption{Quantiles of $\tau \left(\cdot, z, p, q \right)$ as a function of $q$ for $p=0.8, a=20, b=10 \mbox{ and } c=1,000$}
\label{fig:prop2}
\end{figure}

Furthermore, we see in Figure \ref{fig:prop2} that the required number of mutations for $qN$ antibodies to reach the affinity threshold increases approximately linearly with $q$, with a slope that itself increases with the desired confidence level.  In particular, for every affinity landscape $z \in \myel(a,b,c)$ and $0<p,c_1 <1$, there exists a $\beta>0$ such that
$${\frac {R \left(c_1, z, p, q_1 \right) - \beta}{R \left(c_1, z, p, q_2 \right) - \beta}} = {\frac {q_1}{q_2}}.$$
As we would expect from Figure \ref{fig:prop1}, ${\frac {\partial^2 R}{\partial q \partial c_1}} >0$.  Thus, if we want to double the proportion of a GC's antibodies that reach the affinity threshold we have to wait twice as long, while the sensitivity of the response time to the desired confidence level increases as we increase the target proportion of antibodies.

\section{Regularities across affinity landscapes}
\subsection{First two moments}

We now look at how the quantiles of the response time $R$ behave when we vary the affinity landscape.  To effect this, we pick uniformly $m$ independent random sets of $\{a_i\}_{i=1}^m$, $\{b_i\}_{i=1}^m $ and $\{c_i\}_{i=1}^m $ within biologically justifiable ranges as we did in \cite{tedpatty}, and we generate a series of $m$ affinity landscape $z_i \in \myel \left(a_i, b_i, c_i \right)$ respectively.  Let $\mu \left(c_1,p,q \right)$ and $\sigma \left(c_1,p,q \right)$ denote the mean and standard deviation respectively of $R \left(c_1,z,p,q \right)$ over the induced probability distribution on $z$.

\begin{figure}
\epsfxsize=5in
\epsfbox{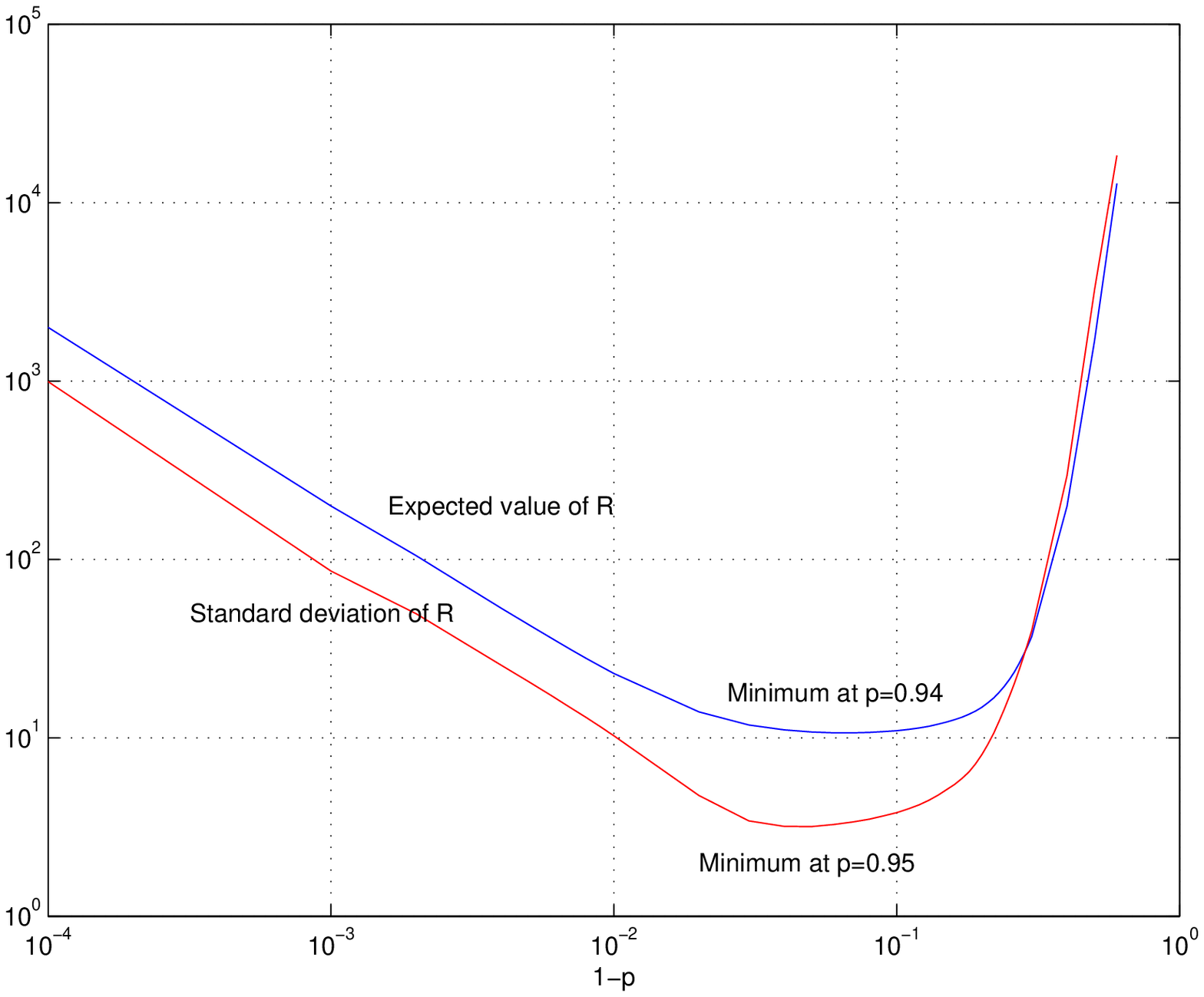}
\caption{The mean and the standard deviation of the response times over 1,000 randomly generated landscapes are almost co-incident ($c_1=50\%$, $q=N^{-1}$ and N=50,000)}
\label{fig:meanstd1}
\end{figure}

Our first observation, shown in Figure \ref{fig:meanstd1}, is that $\mu \left(c_1,p,q \right)$ and $\sigma \left(c_1,p,q \right)$ both exhibit pronounced minima as functions of $p$, and that the corresponding minimizers are almost identical.  Thus, the value of $p$ that guarantees the quickest average response to any antigen whose resulting affinity landscape belongs to $\bigcup_{i=1}^m \myel \left(a_i, b_i, c_i \right)$, also minimizes the variability of the response.  Biologically, the system would likely choose to remain somewhere in this optimal range since this would give a rapid response with high likelihood of reaching threshold affinity even for more difficult antigens, at minimum expense of resources and risk of detrimental mutations accumulating.  This may also permit the greatest number of V gene candidates to attempt to respond to a given antigenic challenge, optimizing search conditions in terms of available pathways and parameters optimized on the landscape such as energetics or kinetics.

\begin{figure}
\epsfxsize=5in
\epsfbox{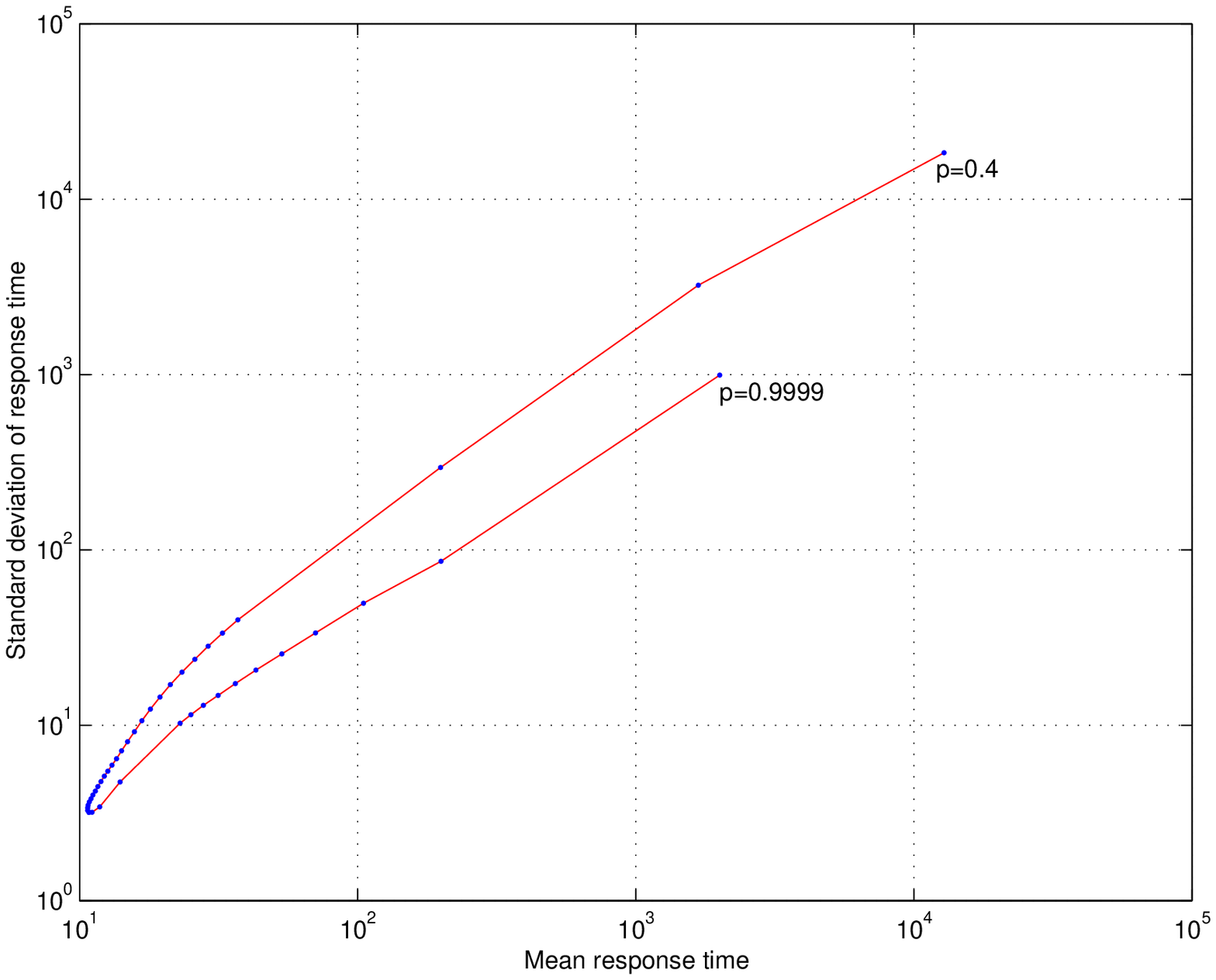}
\caption{Two regimes emerge as we vary $p$ over a large population of affinity landscapes}
\label{fig:cumrobust1}
\end{figure}

\begin{figure}
\epsfxsize=5in
\epsfbox{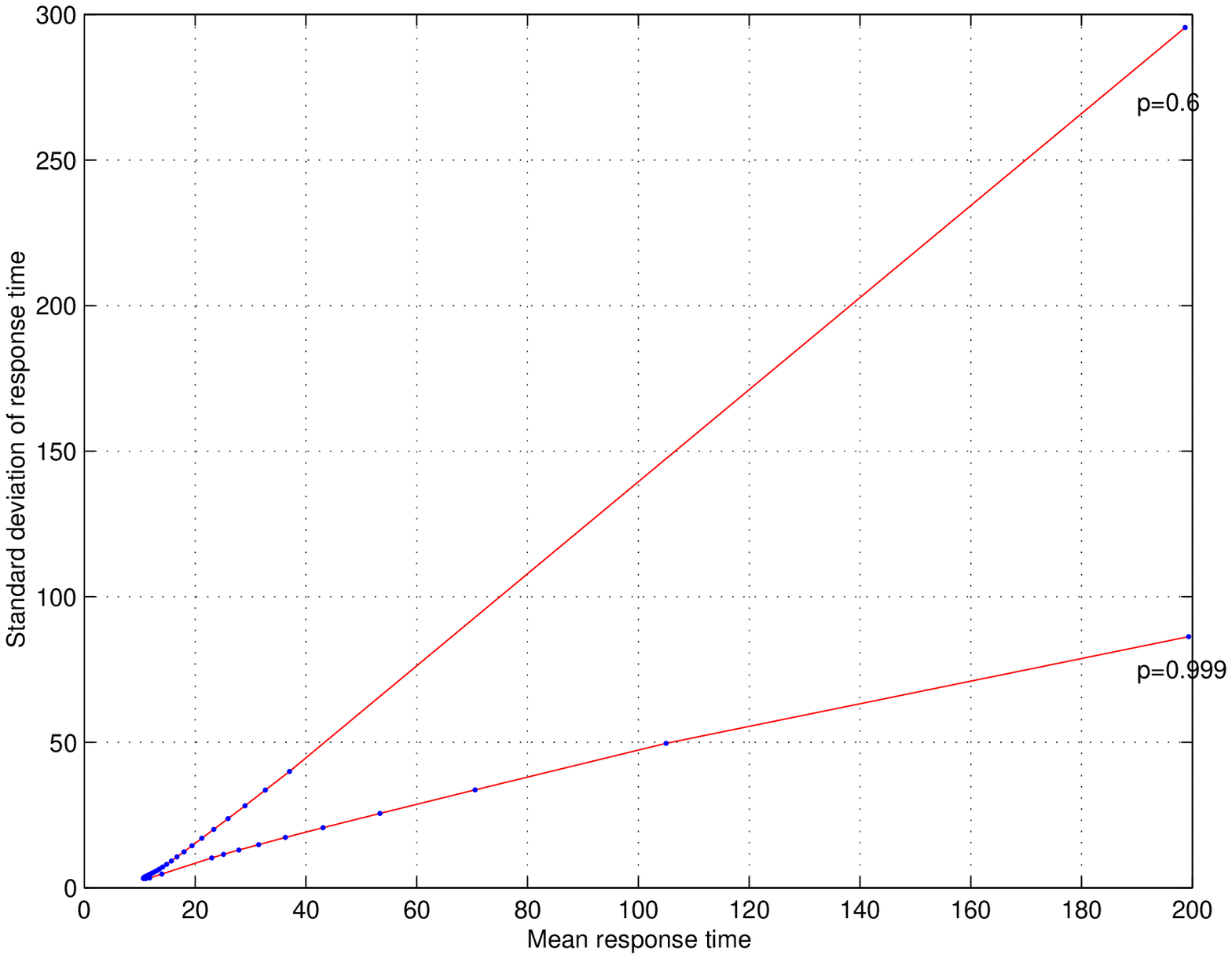}
\caption{Close-up of Figure \ref{fig:cumrobust1} showing the different slopes in the two branches of the graph}
\label{fig:cumrobust2}
\end{figure}

Another way to see this effect is by graphing $\sigma$ as a function of $\mu$ as in Figure \ref{fig:cumrobust1}.  As mentioned earlier in the Note on Random Variables, this figure was generated by iterating our analytic estimates for the desirable response time quantiles as a function of $p$, over different, randomly generated affinity landscapes.  A 3-D graph is first produced, mapping the first two moments of the resulting response time distribution (with respect to the measure of $z$) against the corresponding values of $p$.  Finally, the $p$ dimension is suppressed to focus on the relationship between the first two moments of the response time distribution.  Thus, as labeled on the graph, points on the resulting curve are parameterized by values of $p$.  The same holds for Figure \ref{fig:cumrobust2} which offers a blow-up of the region around the cusp in Figure \ref{fig:cumrobust1}.

Increasing $p$ progressively first leads to a dramatic decrease in both the mean response time and its variability, until we reach a narrow ``critical'' range of $p$ values.  Once we increase $p$ past this range, the response begins to deteriorate, but with measurably lower, yet still increasing response variability. This decreased reliability in the performance of the system in the low $p$ values could due too frequent jumps around the landscape, randomizing the system too frequently.  This does not permit adequate exploration of the local neighborhood. Alternately, $p$ values above the optimal range exhibit a more definite peak, but also show a longer tail.  The system also experiences a decline in performance, perhaps due to getting trapped in local minima without the benefit of a higher level of randomness to escape these minima. Furthermore, Figure \ref{fig:cumrobust2} offers a clear indication that the sensitivity of $\sigma$ to changes in $\mu$ decreases suddenly as we move from the ``low-$p$'' to the ``high-$p$'' regime.

\subsection{Tail behavior}

Our second observation has to do with the shape of ${\rm P}^z \left( R \left(c_1,\cdot, p, q \right) \leq y \right)$ as a function of $y$.  It turns out that for small values of $p$, the pdf of $R$ is heavily skewed, with protracted right tails.  This shape persists qualitatively as we increase $p$ until about $p=0.86$, at which point a transition begins which quickly snaps the right tail and transforms the distribution to approximately uniform.  

\begin{figure}
\epsfxsize=5in
\epsfbox{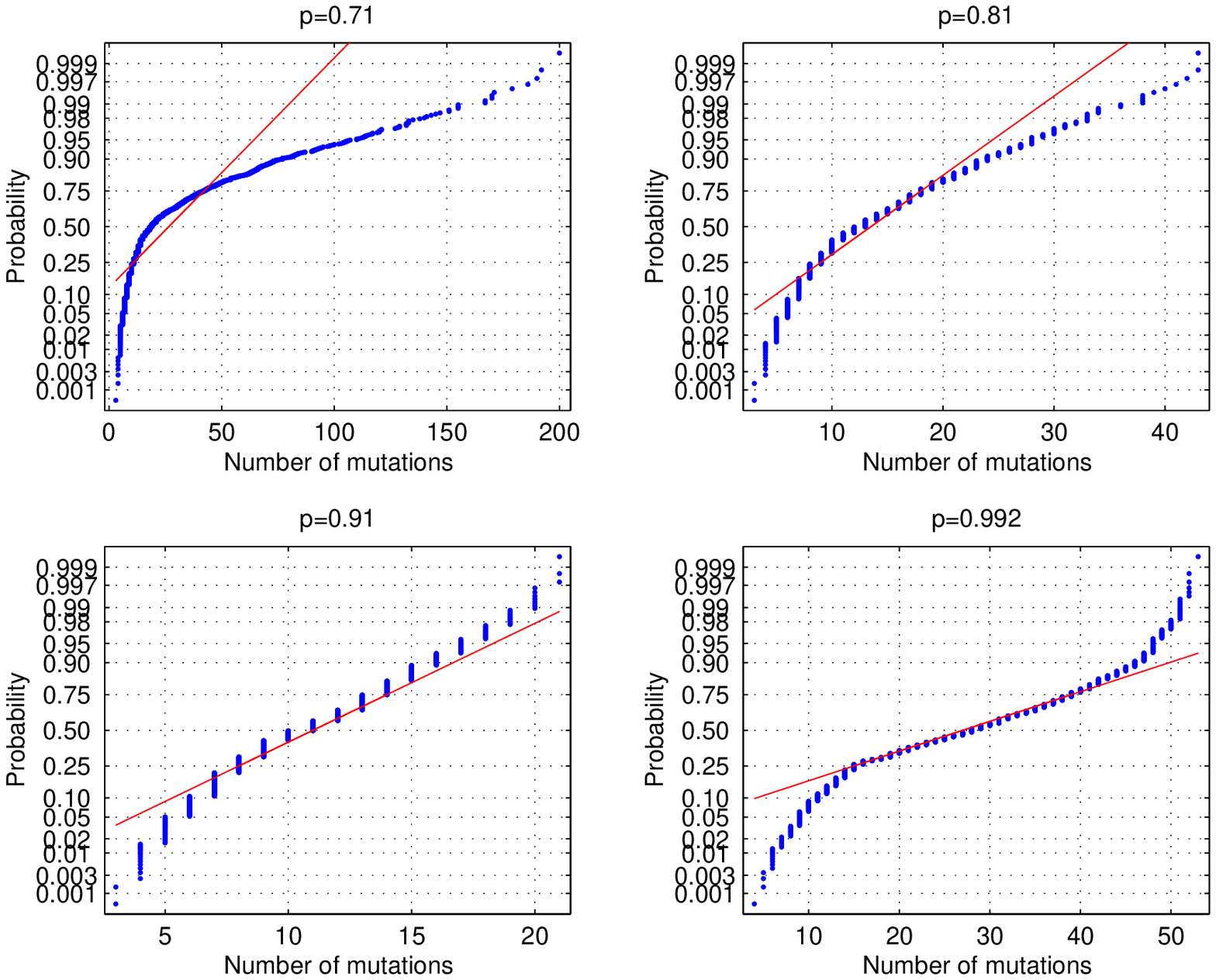}
\caption{As we increase $p$, the tails of $R$ over random affinity landscapes progress from fat to thin relative to the normal (which is represented as a straight line in each graph)}
\label{fig:cumrobust4}
\end{figure}

The normal probability plots in Figure \ref{fig:cumrobust4} compare the tails of $R \left( c_1, \cdot, p, N^{-1} \right)$ against the best-fit normal, i.e. ${\mathcal N} \left(\mu, \sigma^2 \right)$.  These plots clearly show that, while $R$ has fat left tails irrespective of $p$, the right tails die out slower than the normal's ($e^{- {\frac {(x-\mu)^2}{\sigma^2}}}$) for low values of $p$.  As we increase $p$, the right tail is shortened until it looks more like a uniform distribution.

In order to quantify this distributional shift, we computed the Kolmogorov-Smirnov (KS) statistic between $R \left( c_1, \cdot, p_1, N^{-1} \right)$ and $R \left( c_1, \cdot, p_2, N^{-1} \right)$ (computed over 1,000 randomly generated landscapes in $\bigcup_{i=1}^m \myel \left(a_i, b_i, c_i \right)$) while varying $p_1$ and $p_2$ in $(0,1)$.  This procedure generates a matrix $K \left(p_1, p_2 \right)$ which provides us with the confidence levels to which we cannot reject the null hypothesis that $R \left( c_1, \cdot, p_1, N^{-1} \right)$ and $R \left( c_1, \cdot, p_2, N^{-1} \right)$ come from the same distribution.

\begin{figure}
\epsfxsize=5in
\epsfbox{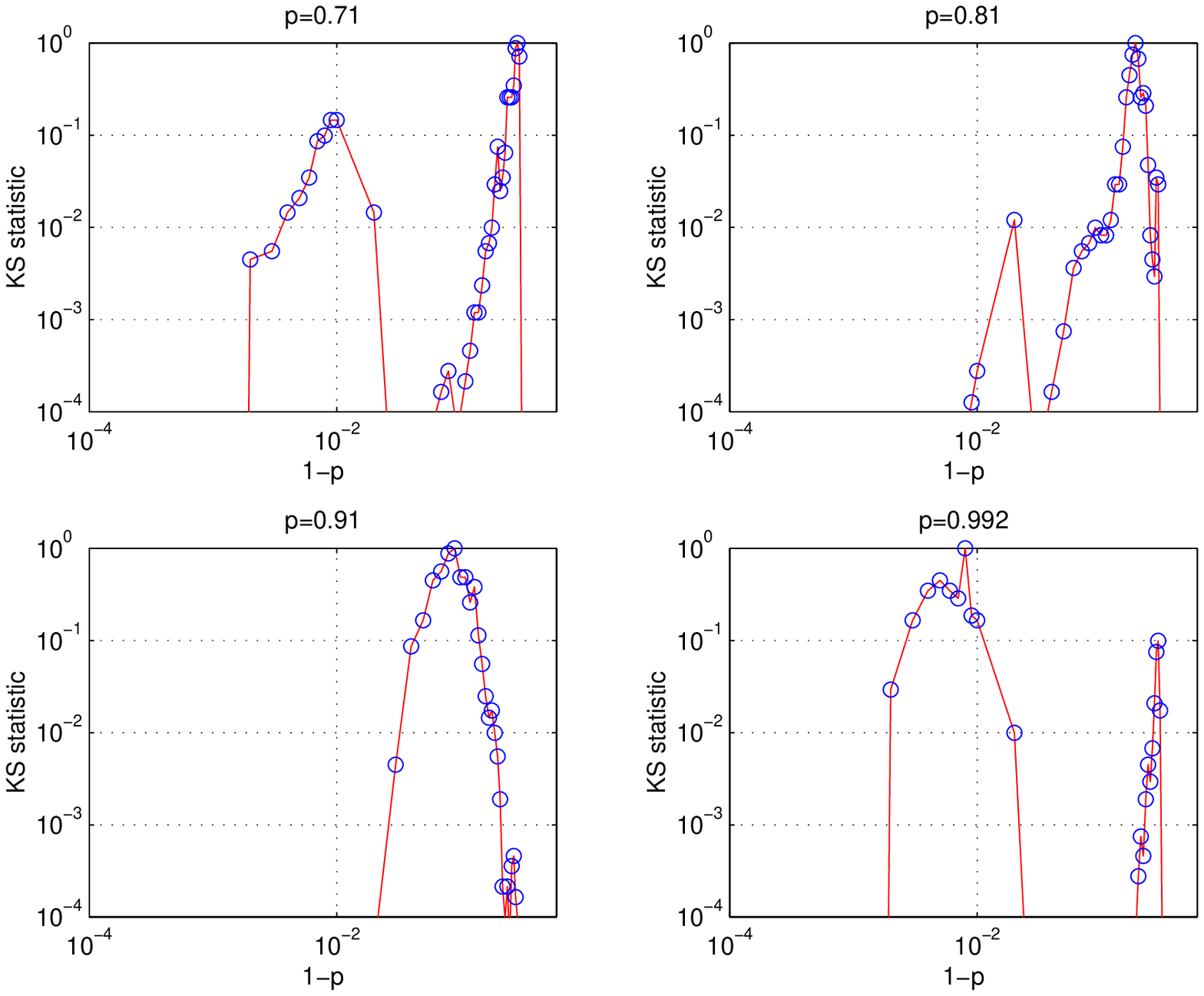}
\caption{Graphs of the KS statistic between $R \left(c_1, \cdot, p_1, N^{-1} \right)$ and $R \left(c_1, \cdot, p_2, N^{-1} \right)$ for $p_1=0.71,\/ p_1=0.81,\/ p_1=0.91 \mbox{ and } p_1=0.992$ and $p_2 \in [0.4,0.9999]$}
\label{fig:cumrobust3}
\end{figure}

Figure \ref{fig:cumrobust3} provides us with four cross-sections of $K$ for increasing values of $p$.  We notice that for low values of $p$, there is more than 10\% chance that $R$'s distributions for all other low values of $p$ are identical.  This phenomenon persists until past $p=0.91$.  However, at some point around $p=0.98$, there is a switch to $R$ distributions arising from high values of $p$, which also have a reasonably high probability of being identical.

\begin{figure}
\epsfxsize=5in
\epsfbox{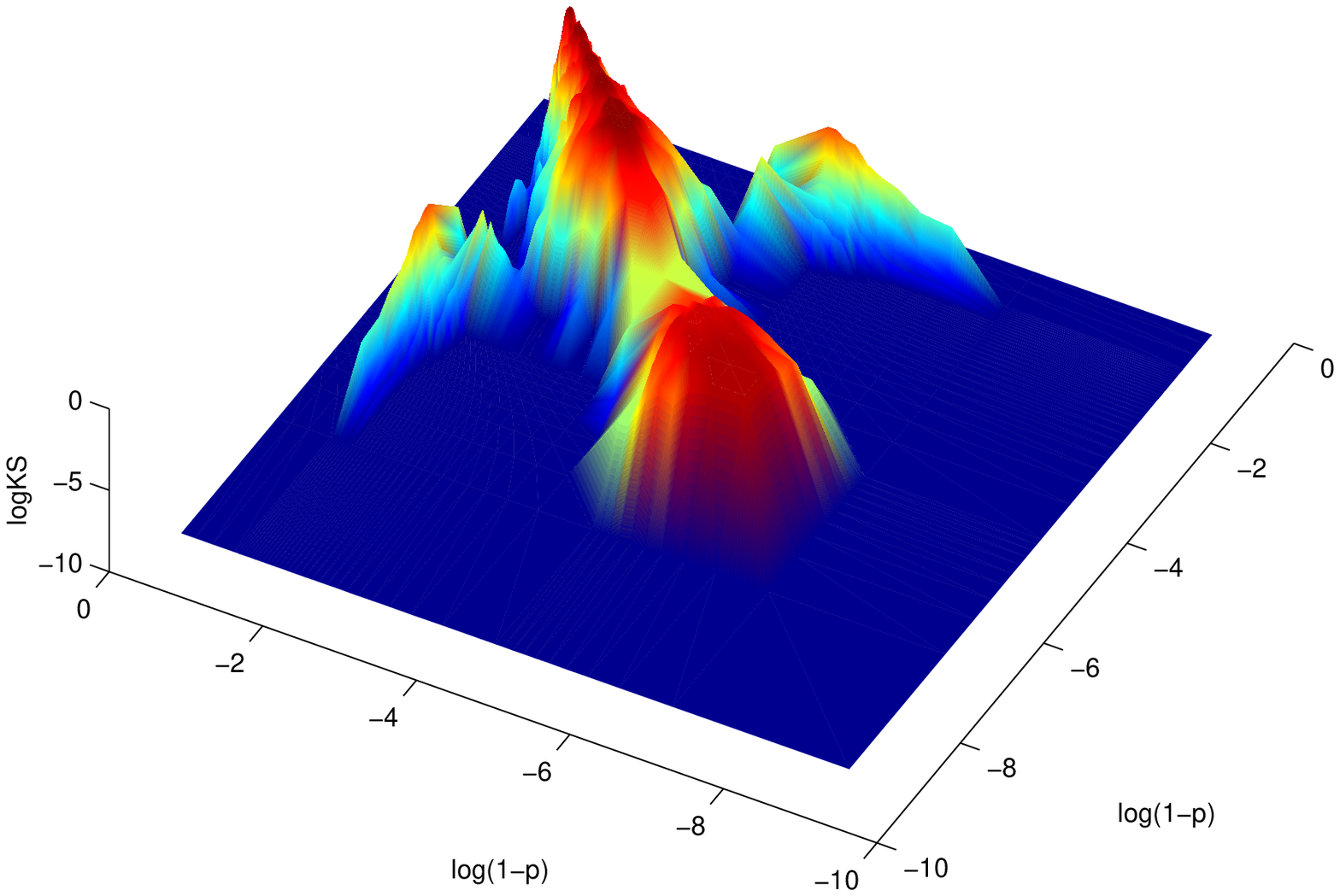}
\caption{The KS statistic between $R \left(c_1, \cdot, p_1, N^{-1} \right)$ and $R \left(c_1, \cdot, p_2, N^{-1} \right)$ for $p_1, p_2 \in [0.4,0.9999]$}
\label{fig:ks11}
\end{figure}

\begin{figure}
\epsfxsize=5in
\epsfbox{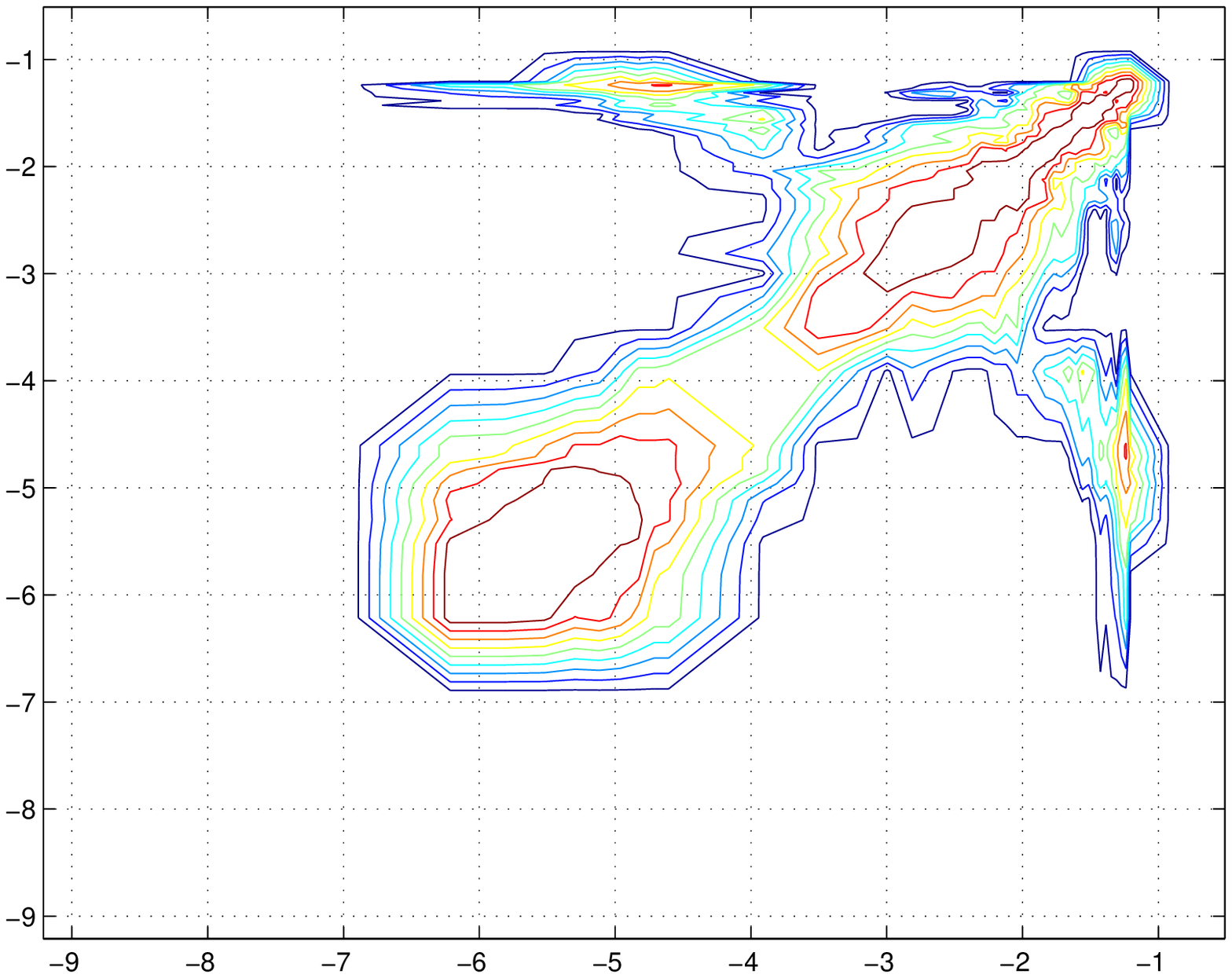}
\caption{Contour plot of Figure \ref{fig:ks11} ($\log(1-p)$ on both axes) showing the two ``islands'' of high distributional stability (one at low values of $p$ and one at high values of $p$), separated by the ``saddle-point'' around $p=0.98$}
\label{fig:ks14}
\end{figure}

This behavior is seen clearly in Figures \ref{fig:ks11} and \ref{fig:ks14}.  Here we see indeed two ``islands of distributional stability'', one for low and one for high values of $p$, separated by a ``saddle-point'', itself very close to the ``turning point'' in the $\mu - \sigma$ graph in Figure \ref{fig:cumrobust2}.

\begin{figure}
\epsfxsize=5in
\epsfbox{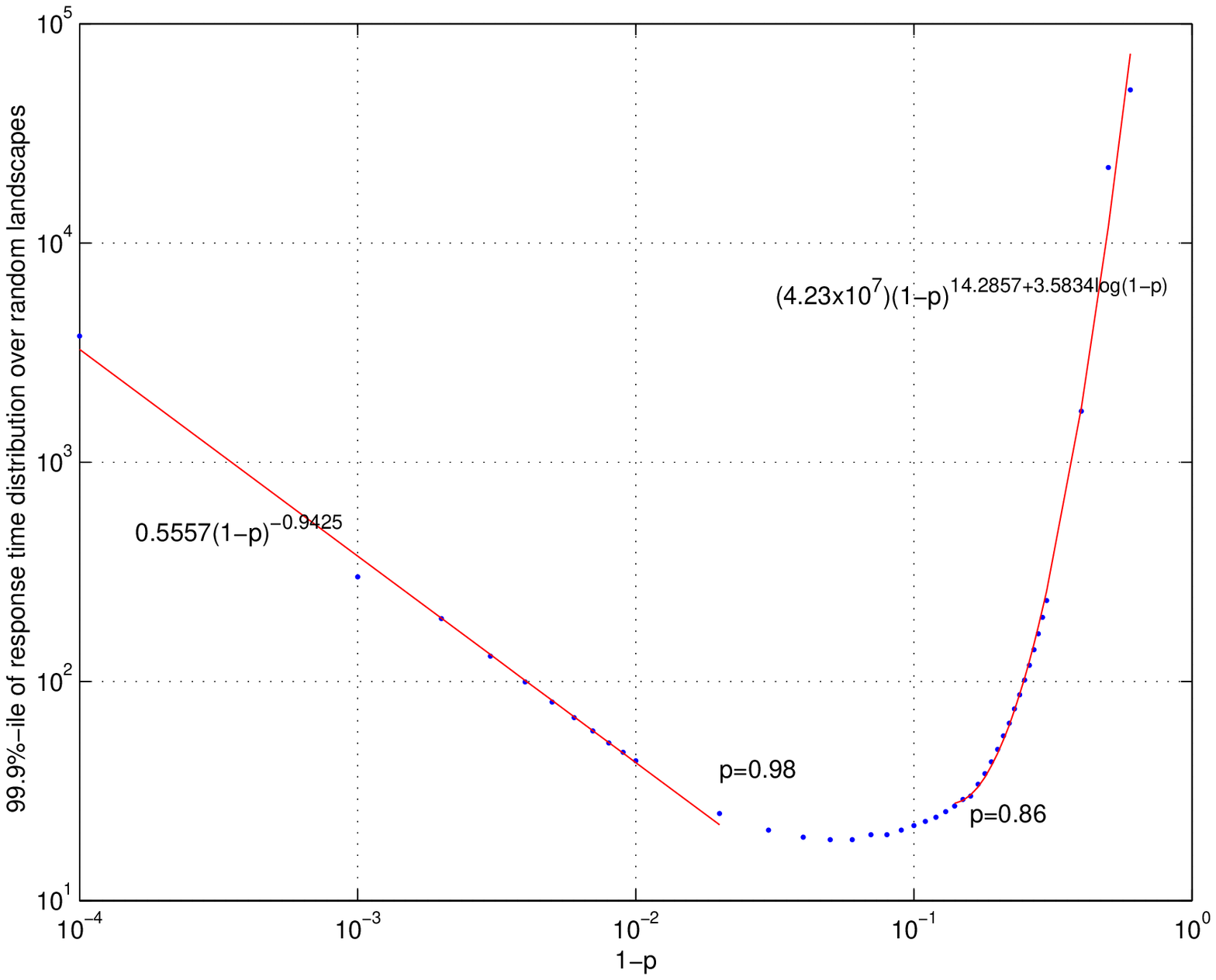}
\caption{Fitting the two branches of the 99.9\%-ile graph for $R$ as a function of $p$}
\label{fig:tailfit1}
\end{figure}

A similar segregation of two regimes is seen in Figure \ref{fig:tailfit1} which shows the different scaling properties of the quantiles of $R$ for values of $p$ below and above a relatively narrow critical range.  Namely, over a large population of affinity landscapes (representing different antigenic agents) the worst-case performance of antibodies exhibiting a $p>0.98$ scales approximately like ${\frac {1}{1-p}}$.  This scaling changes to approximately $(1-p)^{14+4\log(1-p)}$, over a range of $p$ that corresponds to the saddle-point in Figure \ref{fig:ks14}.

In biological terms, the system needs to ensure a consistent response within the constraints of the available number of clones, the need for rapid response and physical and energetic limitations. The lower $p$ value decreases the reliability of the response, and requires the utilization of more energy, resources and time to respond, especially if the antigen is more complex or challenging.  It would also burden the feedback system for screening detrimental mutations.  Alternately, if the $p$ value is too high then again we face a potentially more serious timing problem and risk not achieving threshold affinity in a reasonable time frame. On either side of the optimal range, the risk is present of being unable to mount an adequate response to the antigen and optimally stimulate other immune effector functions.

\section{Evolution of the control parameter $p$}

\begin{figure}
\epsfxsize=5in
\epsfbox{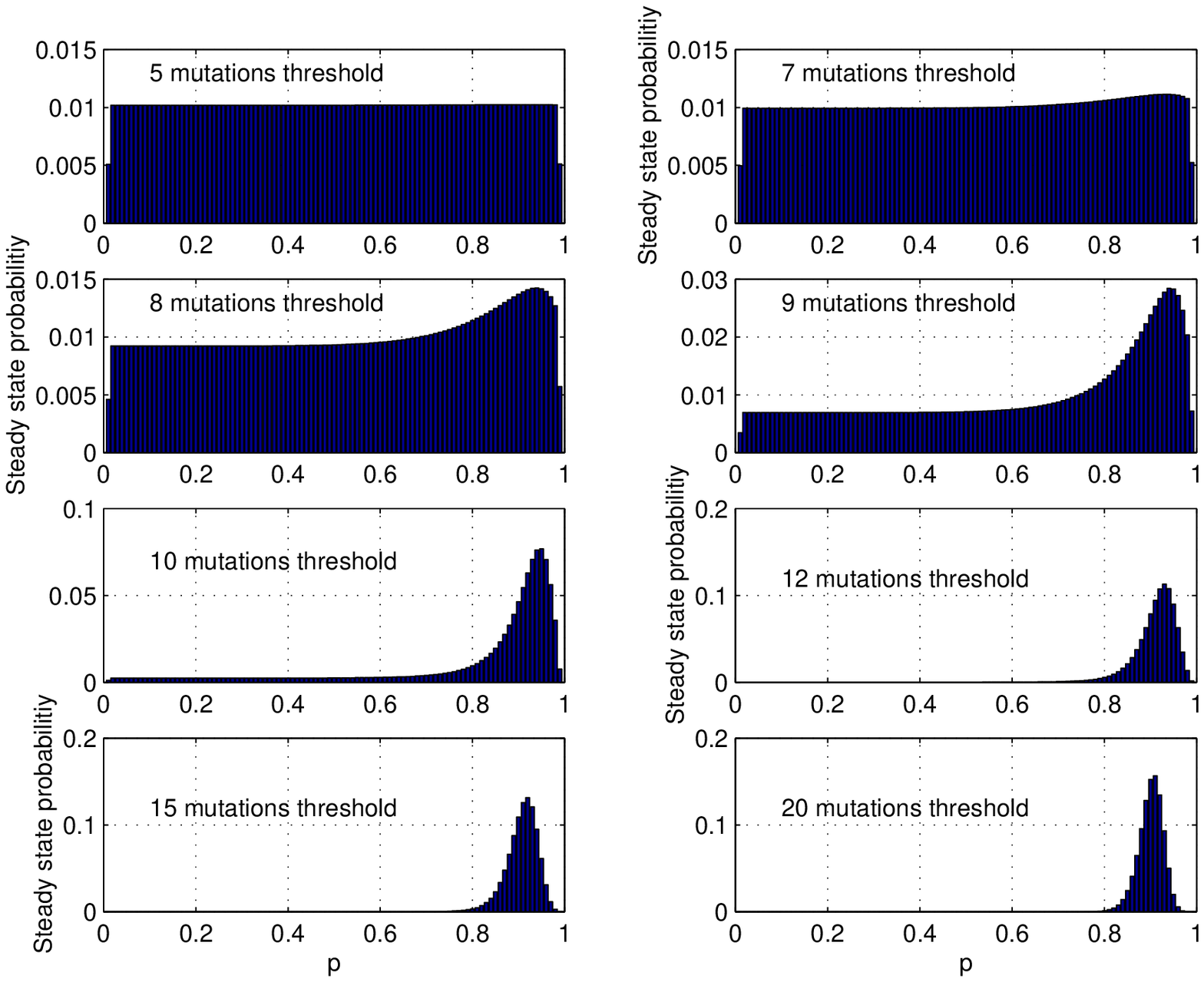}
\caption{Progressive snapshots of the steady-state distribution of $p$, as we decrease the speed of evolution for $p$}
\label{fig:evolve11}
\end{figure}

Previously we used a hierarchical evolution model \cite{tedpatty} to depict the evolving value of $p$ in a population, and the optimal value of $p$ corresponding to the landscape for each new infection. Despite starting distinctly outside the desired range, the system converged rapidly to the 
desired band of $p$ values and was stabilized within that band.  We hypothesized that the performance advantage conferred by this range of $p$ values outweighed the disadvantage of not reaching the optimal $p$ required to respond to any individual antigen. 

In our current treatment of evolvability, we want to better understand how the frequency of mutations in the control parameter $p$ influences its steady-state distribution. We start $p$ randomly, and allow each subsequent mutation to influence $p$ with a small probability.  When the current value of $p$ is far from the value of $p$ that minimizes the response time to the antigen we are facing, then the antibodies will need more mutations, leading to a higher likelihood of an eventual mutation in $p$.  One the other hand, as $p$ approaches the optimal value for the invading antigen, the antibodies respond more quickly, allowing less time for possible mutations to $p$.  

From the above results, we can see that mutating $p$ too frequently leads to insufficient differentiation between different values of $p$ and thus creates a more uniform steady state distribution for $p$; thus too great a variability would likely be detrimental to performance.  At slower mutation rates we see that the distribution begins to peak within the optimal range of values.  This may be evidence of an emerging time limit during the mutation process, where a B cell that has not yet reached threshold by approximately 20 mutations is therefore more likely to mutate $p$ to effect a response.  

From our previous results, we might also predict that the system would choose to maintain $p$ coded in the optimal band, but maintain the ability to mutate it within this band to further optimize performance. Therefore, perhaps any B cell could respond to any antigen, but there may be some best fit advantage conferred to those with closer $p$ values.  It is also possible that there are genomic constraints that keep antigens with response $p$ values in the robust range.  Antigens, though, are myriad and often highly mutable themselves, so we need to allow for the possibility that they can present a range of optimal response $p$. 

If $p$ were coded in the Ig gene itself, then each time there is a mutation in the code, there would be a non-zero probability that $p$ would change as well.  This may be understood by the system as a new likelihood for the probability of the next global jump. The observed mutation distribution during hypermutation is not uniform, because if it were, the predicted degeneracy rate would be too high. Instead it has evolved to target certain sites preferentially 
\cite{manser,dorner,foster,foster2,cowell}. Intrinsic hotspots are usually 
shown to be hotspots independent of antigenic selection, and concentrated in 
the CDRs, often in an overlapping fashion \cite{dorner2,elgert}.  In experiments with a light chain transgene, silent mutation in one part of the gene resulted in loss of a hotspot motif and in the appearance and loss of hotspots in other areas \cite{goyenechea}. This argues for a higher order template as well as an evolving dynamic with loss and acquisition of mutations.  This higher order structure may be conferred by DNA folding, or perhaps DNA--protein interactions \cite{goyenechea}.  Therefore, the optimizing of the mutation frequency of $p$ may also emerge from the V genes themselves, both in their structure and code.

\section{Conclusion}

The system under consideration is complex and dynamic and operates as an orchestra of signals, threshold concentrations, feedback loops and other self regulating behaviors.  It must be responsive to the environment, but avoid extreme sensitivity to initial conditions, so that its robust qualities can be appreciated.  The disequilibrium nature of our performance 
criterion necessitates a computational burden that would make it impractical to exhibit this robustness reliably through simulation 
from first principles.  Our model does not attempt to capture the details, 
or make ad hoc assumptions,  but incorporates biologically sound parameters into 
a simplified framework that yields a deeper result -- the robust and fundamental 
characteristic of the system that is realized in the trade-off represented by 
$p$.  

We can now better appreciate how the organism manages to successfully face 
myriad antigenic challenges.  This ability appears to arise naturally; it is 
not a constant battle for the organism but the natural consequence of its complexity.  Appreciating that antigens and antibodies have co-evolved, each responding to the adaptive pressures of the other, it is also the nature of the antigens that helped shape the evolution of $p$ in this system. 
The information stored in such molecules figures prominently into the naturally arising behaviors we observe.  There is ample evidence for coding bias in the V gene \cite{dorner2,foster2,cowell}, even suggesting that as a result of co-evolution, there exist ``dominant epitopes'' among pathogens with corresponding pre-selected effectors in the response population. In other words, perhaps certain V genes have evolved to be more mutable, and others more robust \cite{oprea2,furukawa,cowell}.  The response and responders are modified accordingly depending on the feedback signals received.  Along with the idea of feedback is the theme of the cells ``sensing'' their environment, both external and internal, and monitoring their performance in order to adjust to these signals.  Our model responds to such a phenomenon- there is no all-or-nothing response, but rather a modulation based on gradients.  
  
While the mutator mechanism is still not understood, certain transcriptional elements appear to be necessary for mutation in both heavy and light chain genes \cite{fukita,klix,diaz,blanden}, although additional, novel molecular mechanisms need to be considered. Evidence that multiple Ig receptors can be expressed on the surface of an individual B cell \cite{kenny} may allow the possibility of such a novel molecular mechanism being employed in affinity maturation as well. A reverse transcription model has already been suggested \cite{blanden} and with recent evidence that mRNA can travel intercellularly, thus altering the phenotype of neighboring cells, makes the possibility of alternate expression paradigms intriguing.  No experimental evidence exists yet to support this in the context of affinity maturation, but our model does not preclude such novel mechanisms.

Finally, the complete appreciation of the significance of $p$ may extend to more global control mechanisms operating throughout the genome, the actions of which may be observed on all scales of evolution.  
  
\bibliographystyle{amsalpha}
\bibliography{immune}

\end{document}